# Far-field e-beam detection of hybrid cavity-plasmonic modes in gold micro-holes

I. Carmeli[1], M. A. Itskovsky[2], Y. Kauffmann[3], Y. Shaked[5],

S. Richter[1]*, T. Maniv[2]* and H. Cohen[4]*

[1] School of Chemistry and Center for nanoscience and nanotechnology, Tel-AvivUniversity, Tel Aviv 69978, ISRAEL

[2] Schulich Faculty of Chemistry, Technion-IIT, 32000 Haifa, ISRAEL

[3] Faculty of Material Engineering, Technion-IIT, 32000 Haifa, ISRAEL

[4] Weizmann Institute of Science, Chemical Research Support, Rehovot76100, ISRAEL

[5] Physics Department, Bar-IlanUniversity, Ramat-Gan52900, ISRAEL

**Manipulation of light-beams with subwavelenth metallic devices has motivated intensive studies, following the discovery of extraordinary transmission of electromagnetic waves through sub-wavelength apertures in thin noble-metal films. The propagation of light in these holes can be investigated at greatly improved spatial resolution by means of focused electron-beams. Here we demonstrate direct e-beam excitation of radiative cavity modes well below the surface plasmon (SP) frequency, of isolated rectangular holes in gold films, illuminating the hotly debated phenomenon of the extraordinary optical transmission through subwavelength holes. The exceptionally long range e-beam interaction with the metal through the vacuum, involving electromagnetic excitations within the light cone, is allowed by momentum conservation breakdown along the e-beam axis. Two types of lowlying excited modes are revealed: radiative cavity modes which are nearly unaffected by SPs, and SP polariton modes with waveguide character in the near field region of the slit walls, which in spite of the strong hybridization preserve the waveguide cutoff frequencies and symmetry characteristics.**

A basic mechanism giving rise to extraordinary optical phenomena in thin metal films[1,2] concerns the breakdown of translational symmetry along sub-wavelength interfaces, which dramatically enhances the coupling between far-field (FF) photons and near-field (NF) surface plasmon polariton (SPP) modes[2,3]. While reasonably understood for arrays of holes[4], and already being exploited for novel applications in, e.g., light-beam shaping devices[5] and biological molecule sensing[6,7], the basic phenomenon of light coupled to SPPs in a *single* hole is still a matter of debate[8,9]. In the heart of the problem is the distribution of the electromagnetic (EM) near-field and far-field across the hole and around its walls. This can effectively be studied by exploiting the superior (~0.1 nm) lateral resolution of a fast (relativistic) focused electron beam (e-beam) at a non-touching configuration[10,11,12,13] inside such a cavity through its spatially scanned electron energy loss (EEL) spectrum.

It has been recently shown[14,15,16] that the corresponding EEL spectral mapping follows the photonic local density of states at the e-beam position, defined in a plane perpendicular to the beam propagation. For a fast e-beam moving in the vacuum near a dielectric medium, excitation of (Cherenkov) radiation has been shown [17,18] to exist via the evanescent tail of the near EM field. Direct excitation from the vacuum FF zone of EM *radiation* by the e-beam, involving scattering channels inside the light cone (i.e. under momentum conservation breakdown along the e-beam direction), has been predicted theoretically[19], but so far has not been verified experimentally. In this article we present for the first time clear experimental evidences for this striking feature of a fast e-beam, which enables to detect the usually missing radiative components of the EM field in the entire space of a cavity. We have found that the e-beam interacts with the metal through the vacuum at exceptionally long distances (~400nm) by exchanging photons within the light cone[19], revealing low energy waveguide (WG) resonances in the EEL spectrum well bellow the main surface plasmon (SP) peak-position.

Two types of WG modes are observed: radiative modes which are nearly unaffected by surface plasma oscillations, and hybrid modes of SPP propagating on the hole boundaries with WG character in the near field region of the slit walls. We show that in spite of the strong hybridization, the hybrid modes preserve the waveguide cutoff frequencies and symmetry characteristics.

For the sake of demonstration of the main findings we focus in the present work on a relatively large (i.e. micrometer wide) hole, and present only briefly our results for a submicron-cavity.

Fig. 1a shows a set of spectra, scanning the long principal axis (y) of a 900x4500 nm$^2$ slit in a free-standing gold film 200 nm thick. The appearance of low-energy signals can be easily noticed; their details better seen after background subtraction, as shown in Fig. 1b, main panel. A major advantage of this spectroscopy mode over energy filtered images [20] regards the background subtraction, which turns out in this system to be a critical step (see Methods). Significantly below the gold SP at 2.4 eV, peaks are observed at 0.65 eV and ~1.9 eV and, with smaller intensity, also around ~ 1.0 and 1.3 eV. The (seemingly noisy) spectrum shows consistently these low-energy peaks in large sets of data. Its 0.65 eV signal is attributed to a standing wave dictated by the 900 nm slit width (in the x-direction). A similar set of spectra is shown in Fig.1c for a 600x4500 nm$^2$ slit drilled in the same gold film, showing a pronounced low energy signal at about 1 eV, which is attributted to a standing wave dictated by the 600 nm slit width.

As support to these assignments, spectra recorded from different slits are compared, Fig. 2a, showing the expected reciprocity between the slit width and the low-energy peak position. Moreover, integer multiples of the fundamental frequency can be observed in these spectra: the second harmonic lines in Fig. 2a are relatively weak, but the third harmonic peaks are again more intense. Table 1 compares the experimentally derived peak positions with theoretical energies calculated for WG resonances of infinitely long rectangular slits: $\omega_m \approx c(\pi m / a)(1-\delta)$, $m = 1, 2, 3, ...$, where $a$ is the slit width and $\delta \sim 0.05 - 0.1$ is a correction due to deviation from the perfect conductor (PC) boundary conditions (see below). The agreement found between calculated and measured peak positions strongly supports their assignement to WG resonances.

The appearance of EELS signals at distances larger than 400 nm from any of the slit edges is far beyond the distances associated with typical NF evanescent tails. It demonstrates for the first time that FF interactions from the vacuum, involving excitations within the light cone, can be realized in EELS, implying that the e-beam can interact with the metal through the vaccum at exceptionally large distances due to breakdown of momentum conservation along the e-beam axis[19]. As usually expected, the intensity plots presented in Fig. 2b show rapid signal decay from the slit edge towards the vacuum at distances of the order of 100 nm, reflecting the essentially NF nature of the interaction in this region. However, the low energy signals (except for the second harmonic) remain clearly observable at much larger distances, revealing the residual FF component of the (nearly distance-independent) interaction.

To gain deeper insight into this remarkable phenomenon we consider here a simple model of a highly focused e-beam propagated within a slit in non-touching paths at ascending distances from its walls. The slit is assumed to be infinitely long in the y-direction and of width $a$ along the x-axis. The corresponding EEL spectra are calculated by employing an extended version[19] of the method developed in Ref.21: The effective range of the force acting on the e-beam is restricted to a finite length of the order of the film thickness, $2L$, while keeping the form of the self-induced EM interaction the same as in an infinite slit. The model calculation, which accounts for multiple reflections between the parallel slit walls, yields for the differential loss probability:

$$\frac{d^2P}{d\omega dz} = \left(\frac{e}{4\pi^2\hbar\omega L}\right) \text{Re}\left[\int_0^\infty dk_z \rho_L(k_z - \omega/v) \int_0^\infty dk_y E_z(x_e, k_y, k_z, \omega)\right] \quad (1)$$

where $\rho_L(k) \equiv \sin(kL)/k$, $v$ is the electron velocity along the beam direction, and the expression for the 2D Fourier transform (with wave numbers $k_y, k_z$) of the electric field component involved in the loss process at the position of the e-beam with respect to the slit center, $x_e = x - a/2$, is:

$$E_z(x_e, k_y, k_z, \omega) = A_e \rho_L(k_z - \omega/v) \left[ \begin{array}{c} ik_z \chi_1 \left( \tilde{F}_+ e^{\chi_1 x_e} + \tilde{F}_- e^{-\chi_1 x_e} \right) \\ + \left( \left(\frac{\omega}{c}\right)^2 - k_z^2 \right) \left( 1 + \tilde{A}_+ e^{\chi_1 x_e} + \tilde{A}_- e^{-\chi_1 x_e} \right) \end{array} \right] \quad (2)$$

$$\text{with: } \chi_1 \equiv \sqrt{k_z^2 + k_y^2 - \left(\frac{\omega}{c}\right)^2}, \quad A_e \equiv \pi e / i\omega \chi_1,$$

$$\tilde{A}_\pm \equiv \frac{A_\pm}{A_e} = \frac{(1 - \chi_2/\chi_1)}{2} e^{-\chi_1 a/2} \left\{ \begin{array}{c} \dfrac{\cosh(\chi_1 x_e)}{\sinh(\frac{\chi_1 a}{2}) + (\chi_2/\chi_1)\cosh(\frac{\chi_1 a}{2})} \\ \pm \\ \dfrac{\sinh(\chi_1 x_e)}{\cosh(\frac{\chi_1 a}{2}) + (\chi_2/\chi_1)\sinh(\frac{\chi_1 a}{2})} \end{array} \right\} \quad (3)$$

and:

$$\tilde{F}_{\pm} \equiv \frac{F_{\pm}}{A_e} = \pm\left(\frac{ik_z}{2\chi_1}\right)(1-\varepsilon(\omega)) \times$$

$$\left\{ \frac{\cosh(\chi_1 x_e)}{\left[\varepsilon(\omega)\cosh(\frac{\chi_1 a}{2}) + \left(\frac{\chi_2}{\chi_1}\right)\sinh(\frac{\chi_1 a}{2})\right]\left[\sinh(\frac{\chi_1 a}{2}) + \left(\frac{\chi_2}{\chi_1}\right)\cosh(\frac{\chi_1 a}{2})\right]} \pm \frac{\sinh(\chi_1 x_e)}{\left[\varepsilon(\omega)\sinh(\frac{\chi_1 a}{2}) + \left(\frac{\chi_2}{\chi_1}\right)\cosh(\frac{\chi_1 a}{2})\right]\left[\cosh(\frac{\chi_1 a}{2}) + \left(\frac{\chi_2}{\chi_1}\right)\sinh(\frac{\chi_1 a}{2})\right]} \right\} \quad (4)$$

Here $\chi_2 \equiv \sqrt{k_y^2 + k_z^2 - \varepsilon(\omega)\left(\frac{\omega}{c}\right)^2}$, and $\varepsilon(\omega)$ is the metal dielectric constant. The distribution function $\rho_L(k_z - \omega/v)$, which arises from the electrical current density associated with the e-beam, takes into account the finite path of active interaction with the metal. In the limit of an infinitely long cavity in the beam direction, $L \to \infty$, $\rho_L(k_z - \omega/v)$ collapses to a delta function at $k_z = \omega/v$. However, for finite $L$, the momentum conservation condition $k_z = \omega/v$ is not strictly satisfied and small values of $k_y$ and $k_z$ below $\omega/v$ become accessible, allowing wavenumbers, $K \equiv \sqrt{k_z^2 + k_y^2} < \left(\frac{\omega}{c}\right)$, within the light-cone (i.e. with purely imaginary $\chi_1$) to reach resonant conditions of the coefficient $F_{\pm}$ (see Eqs.2,4).

Fig.3a shows the calculated loss probability, based on Eqs.1-4, for a 300 keV e-beam inside a slit of width $a = 900$ nm at a distance (impact parameter) 400 nm from a slit edge for various values of film thickness, $2L$. At this impact parameter the gold SP resonance at 2.4 eV is fully suppressed due to its mainly nonradiative nature. On the other hand, radiative modes like the first two odd WG resonances, with thresholds at 0.65 and 1.96 eV given in Table 1, domiante the spectra. The calculated EEL spectra, shown in Fig.3a, also demonstrate robustnessof the fundamental WG resonance with respect to an increase in the film thickness parameter $2L$. The sharp intensity suppression of the calculated harmonics under increasing values of $L$, in contrast to the fundamental standing wave, may account for the fact that they usually appear weak in the experimental data.

The cutoff frequencies in Table 1 deviate from the corresponding ideal WG frequencies [22] due to deviations from the PC conditions at the slit boundaries, $x_e = \pm a/2$. They were calculated from the resonant condition of $F_{\pm}$ in Eq.4 for $k_y, k_z \to 0$, which yields: $\omega_m \approx c(\pi m/a)(1-\delta), \delta = 2c/a\omega_p, m = 1,2,3,...$, with $\omega_p$ being a bulk plasmon frequency parameter. The corresponding (self-induced at the e-beam position) electric field component has the form $E_z^{odd}(x_e) \propto \cos^2[(\pi m/a)(1-\delta)x_e]$ for odd values of m, and $E_z^{even}(x_e) \propto \sin^2[(\pi m/a)(1-\delta)x_e]$ for the even ones (see Fig.3b). Thus, despite apparent deviation from the PC conditions, the even-mode fields should vanish at the slit center, $x_e = 0$, a prediction which is essentially confirmed by the weak appearances of the corresponding signals in the experiment. Their presence, however weak, particularly that of the $m=2$ mode in the FF region shown in Fig.1c, is attributed to local symmetry breaking associated with structural imperfections.

An interesting finding arising from our experimental data is the apparently selective NF enhancement of low energy peaks with respect to the different slit edges: the enhancement is restricted to those edges that are parallel to the direction of the standing-wave (the x-axis). Fig. 4, bottom panel, presents spectra taken at 8 nm from each of the two perpendicular walls of a 900x4500 nm$^2$ slit. Enhancement of the fundamental and third harmonics appears selectively near the short edge, i.e. the wall parallel to the (900 nm related) standing wave direction. With a smaller slit, 180x900 nm$^2$, one finds similar NF enhancement near the long edge only, i.e. again the edge parallel to the associated standing waves. Note that, unlike the low energy peaks, the dominant NF appearance of the main SP resonance at 2.4 eV does not show any selectivity with respect to the different slit edges.

These results contrast our expectation from EEL signal for an infinitely long rectangular WG at the lowlying cutoff energies since the corresponding electric field component, $E_z$, that couples effectively to the e-beam, is strongly suppressed near the $y=0,b$ walls due to the large (negative) real part of the metal dielectric constnat in the corresponding frequency range. As the EEL peak-positions in the NF region retain their corresponding energies in the FF region, we attribute them to hybride WG-SPP modes consisting of SPPs propagating in the x-z directions of the $y=0,b$ slit walls, which are trapped between the $x=0,a$ walls. The resulting standing waves generated along the x-axis have resonant WG features at the appropriate cutoff frequencies, whereas the termination of the slit walls at the outer faces of the thin metal film provides the required large $E_z$ component through mixing with SPPs propagating on top and bottom of the film faces[9].

Thus, the relatively intense NF EEL peaks observed for the odd $m$ modes arise from such hybride SPP-WG modes. The relatively short range ($\sim 100 \text{ nm} \ll 900 \text{ nm}$) of the NF interaction, seen in Fig.2b, is explained as arising from the sharp spatial rotation, as a function of y, of the electric field vector around the rims of the slit [9]. A large portion of the e-beam path inside the slit is exposed to this rotation due to the small (200 nm) gold film thickness.

Hybridization of lowlying WG modes with surface plasmons also accounts for the observed EEL signals in the case of smaller (e.g. the 900x180 nm$^2$ in Fig. 4) slits, though apparent differences arise due to the small width along the y-axis. Here the low-cutoff TM energy ( TM(1,1) ) of an ideal 2D rectanguar slit with $b = 180$ nm should be considerably higher than the observed low-energy peaks (see e.g. Fig.4a), whereas the lowlying TE(1,0) WG mode, with energy that fits the experimental data, can only weakly couple to the e-beam due to its null $E_z$ field component. However, as shown recently by Gordon and Brolo[23], SP oscillations on the wide slit walls, which are strongly coupled via their slowly decaying evanscent tails, drastically modify the low energy WG activity, introducing a new lowlying TM-like mode with a significant $E_z$ field component near the wide slit walls.

In summary, direct radiative e-beam excitation of WG-like modes propagating in an isolated slit in a thin gold film was observed at extremely large distances of the e-beam from any of the slit edges. The corresponding long range e-beam metal interaction, involving EM excitation within the light cone, is allowed by momentum conservation breakdown along the e-beam axis, revealing WG resonanceswith energies well bellow the SP resonance of the metal. The focused e-beam offers suprior scanning resolution across FF and NF regimes of the hole, providing information on both the WG and the surrounding metal-supported SPP excitations. Our NF results show pronounced hybridization between cavity and plasmonic modes that, remarkably, retain the energy and major symmetry characteristics of ideal WG modes. These findings open a promising new direction in the emerging field of photonics-plasmonics at subwavelenth scales, which could, in particular, shed light on the phenomenon of extra-ordinary optical transmissionthrough sub wavelength apertures in metal films.

## Methods

**Sample preparation**: Free standing Au films were prepared by evaporating 200 nm Au on a NaCl crystal (Agar Scientific), subsequently cut to ~3mm$^2$ pieces and placed in a dish of deionized water until the gold leaf was detached and floated on the water surface. The leaf was then mounted on an Au TEM grid (SPI supply) and left under dry air. Slits of different dimensions were drilled with a focused Ion Beam (FIB). Before insertion to the electron microscope, samples were plasma cleaned for 5 min.

**EELS measurements** were performed on FEI Titan transmission electron microscope (TEM), using a 300 keV monochromated beam, about 5 A in diameter, in the scanning mode. The elastic peak (EP) FWHM ranged between 150-200 meV. Part of the energy broadening arose from relatively slow fluctuations that could be eliminated by using short exposures, 10 ms each, repeated 20-30 times at each pixel. Those exposures that monitored a clear energy jump (with a drastically distorted EP) were rejected. The EP position was then aligned via offline processing, such that the spectra could be reliably summed, giving improved statistics with minimal EP-tail intensity.

For reliable background (BG) subtraction, a reference spectrum was recorded away from the gold film, providing a record of the (instrumental) EP shape. A smooth residual BG was usually left after subtraction of this reference, indicating the presence of low-energy continuum in the excitation spectrum, to be discussed elsehere. For quantitative analyses, this residual BG was removed by a smooth bi-exponential function, fitted to the spectrum in the range of 0.3-6.0 eV, followed by curve fitting analysis for peak intensity evaluation using the Casa softaware.With these experimental procedures, reliable quantification of signals above ~0.5 eV could be achieved.

# Acknowledgements

This work was supported by the Russell Berrie Nanotechnology Institute at the Technion-Israel Institute of Technology (TM), and by James Frank and Wolfson Foundations at Tel-Aviv University (SR)

# Author contributions

I.C. initiated the study, designed the samples, carried out STEM-EELS measurements, performed the data analysis, and was involved in writing the paper. M.A.I participated in developing the theoretical model and carried out the calculations, Y.K. carried out the STEM-EELS measurements, Y.S. performed the FIB operation, S.R. supervised parts of the experimental effort and contributed to the data analysis, T. M. led the theoretical and data interpretation efforts and wrote the paper, H. C. led the experimental work and the data analysis, assisted the theoretical data interpretation, and wrote the paper.

**Figure Captions**

**Figure 1**: A line scan recorded along the long axis of the 900x4500 nm² slit shown in inset I: (a) before and (b) after background subtraction. The black arrows indicate peak positions of the main resonances. The white arrow in inset I (panel a) indicates the scan path. Inset II (panel b) exemplifies the spectral shape after subtracting the reference spectrum (before the second stage of background subtration). Spectra in b are shifted vertically for visual convenience. Beam positions (2250, 1000, 450, 300, 200, 100, 80, 60, 50, 40, 32, 24, 16, 8, and 0 nm, measured from the slit edge) are indicated. Note the appearance of low-energy peaks in addition to the surface plasmon at 2.4 eV. (c) A line scan along the y-axis of a 600x4500 nm² slit after backgound subtraction, similar to that shown in Fig.1b. Note the higher energies of the main resonances (compared to the 900x4500 nm² slit) appearing for smaller slit widths. The inset in panel c presents the original spectra (before backgound subtraction).

**Figure 2**: (a) Spectra recorded at 8 nm from the narrow edge of different slits, all being 4500 nm long, their widths: 600, 900 and 1200 nm. As expected, the peaks shift to higher energies under decreasing slit width. (b) Intensity plots for selected signals in the 900x4500 nm² slit of Fig. 1b demonstrating the fast decay at the NF regime and the FF interaction which takes over at long distances. The experimental error in these intensity plots is ≤30 cps.

**Figure 3**: (a) Calculated EEL spectra of a beam positioned 400 nm from the edge of a slit 900 nm wide and infinitely long, for different values of the gold film thickness (2L). Resonances with thresholds at 0.65, 1.3, and 1.95eV, corresponding to the WG cutoff frequencies $\omega_m \approx c(\pi m/a)(1-2c/a\omega_p)$ with $m=1,2,3$ respectively, are seen. Note that the lowest even WG resonance with threshold at 1.3 eV appears significantly weaker than the other, odd WG peaks due to the e-beam proximity (50 nm) to the slit center (see Fig.3b). Also note the sharp suppression of the high-harmonic resonances for increasing film thickness, beyond 200 nm, and the robustness of the fundamental resonance for thicknesses up to more than 1000 nm.

(b) Dependence of the calculated jumps in the EEL signal at the thresholds $m=1,2,3$ (JL1, JL2 and JL3, representing the large symbols: circles, squares and triangles, respectively) on the beam position $x_e = x - a/2$ (i.e. measured with respect to the slit center). Also plotted, for comparison, are the square moduli of the corresponding ideal WG functions [22] (WG1, WG2 and WG3, representing the small symbols: circles, squares and triangles, respectively). The latter were normalized by adjusting their maximal values to those of the former. Note that significant deviation from the PC boundary conditions takes place only at the third cutoff frequecy (1.95 eV), which is rather close to the surface plasmon frequency at 2.4 eV.

**Figure 4**: Comparison between spectra at 8 nm from the narrow (N, red) and wide (W, black) edges of a large, 900 x 4500 nm² (bottom panel), and of a small, 180 x 900 nm² (top panel) slit, showing enhanced intensities near the 900 nm wide slit wall in both cases.

**Table 1**

Calculated cutoff energies and and measured peak positions for the investigated slits

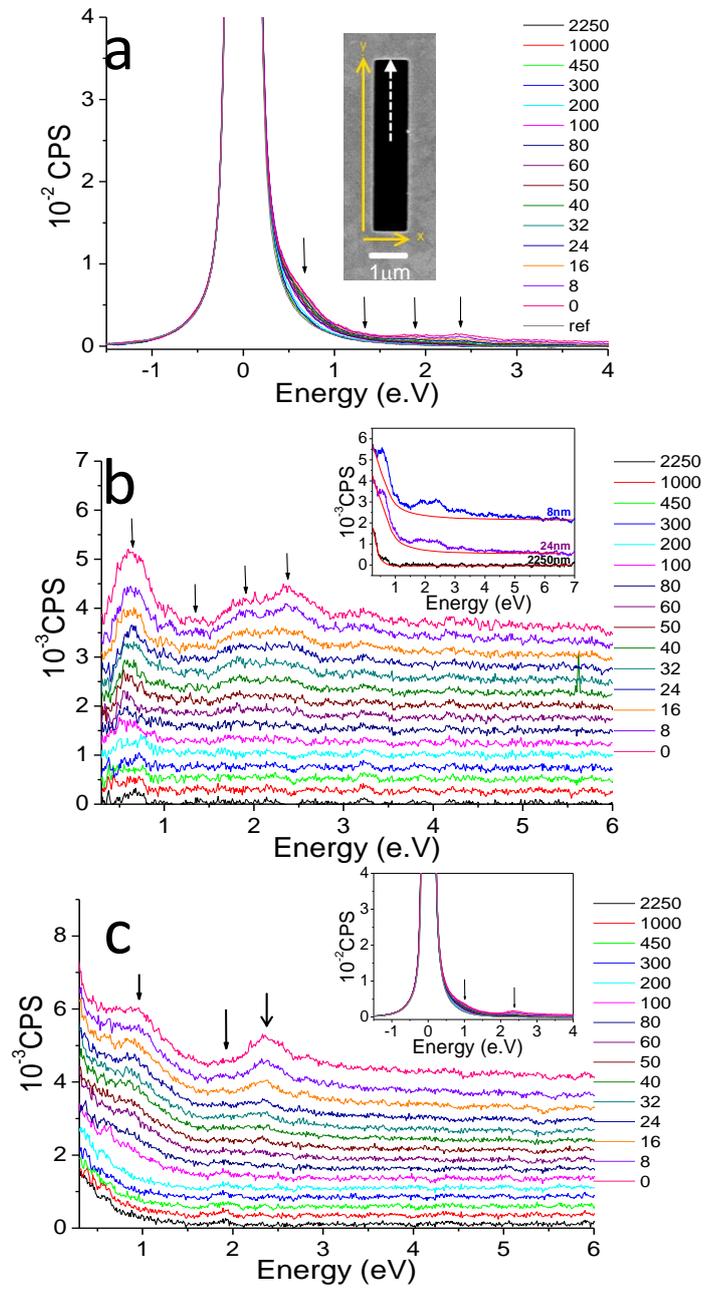

**Fig. 1**

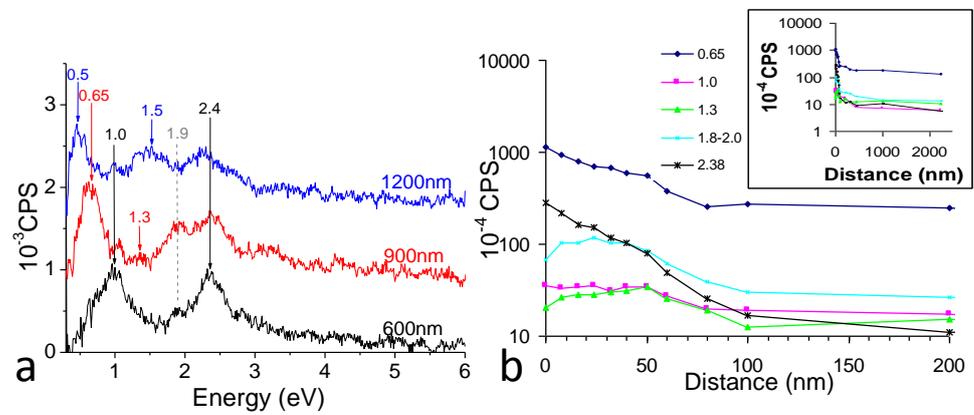

**Fig. 2**

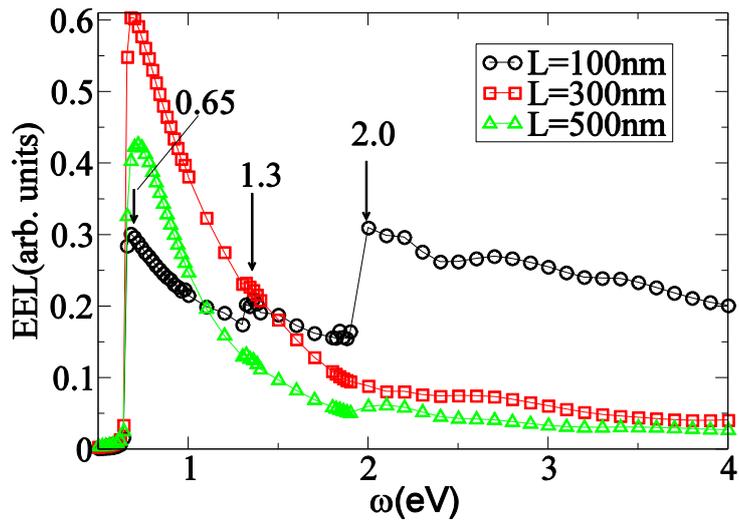

a

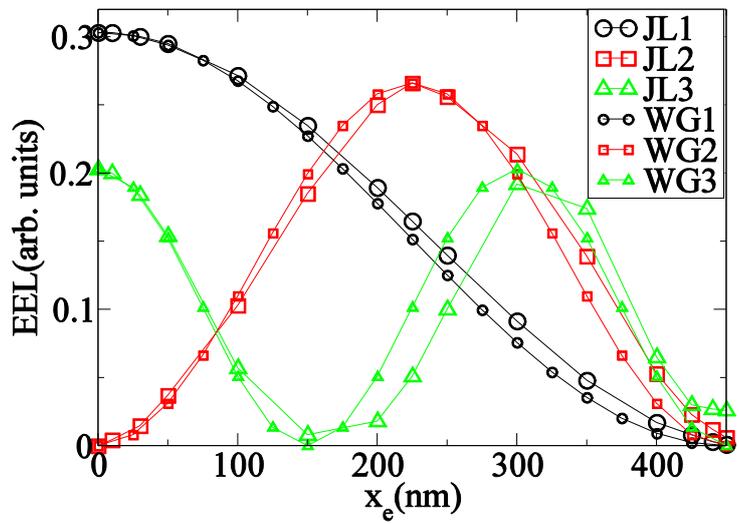

b

**Fig. 3**

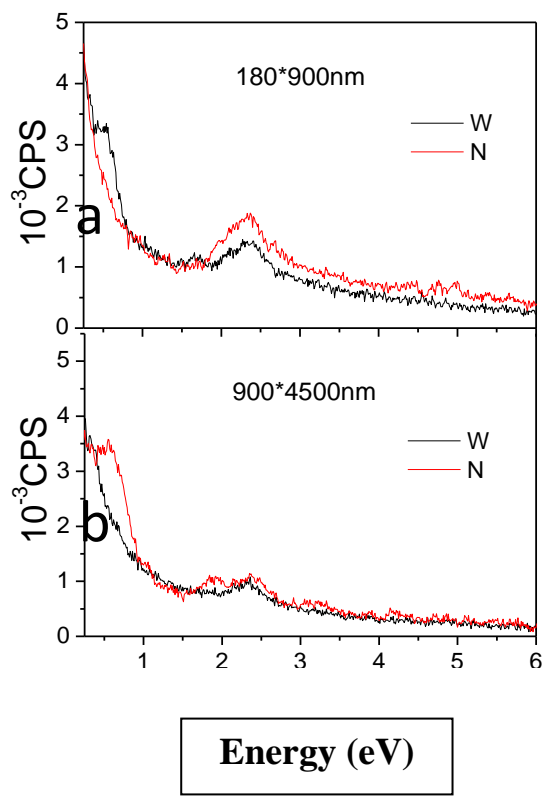

**Fig. 4**

| a (nm) | $\hbar\omega_1$ (eV) | $\hbar\omega_2$ (eV) | $\hbar\omega_3$ (eV) | $\hbar\omega_4$ (eV) |
|---|---|---|---|---|
| 900 | | | | |
| Calculated | 0.65 | 1.3 | 1.96 | |
| Measured | 0.65 | 1.3 | 1.9 | |
| 600 | | | | |
| Calculated | 0.98 | 1.96 | | |
| Measured | 0.95 | 1.9 | | |
| 1200 | | | | |
| Calculated | 0.49 | 0.98 | 1.46 | 1.96 |
| Measured | 0.5 | 1.0 | 1.5 | |

**Table 1**